\documentclass[prl,twocolumn]{revtex4}
\usepackage{amssymb}

\usepackage{graphicx}
\usepackage{multirow}

\newcommand\beq{\begin{equation}}
\newcommand\eeq{\end{equation}}
\newcommand\bea{\begin{eqnarray}}
\newcommand\eea{\end{eqnarray}}

\begin{document}

\title{
The modulated spin liquid: a new paradigm for URu$_2$Si$_2$}
\date{\today}
\author{ C. P\'epin$^{1, 2}$, M. R. Norman$^{3}$, S. Burdin$^{4}$ and A. Ferraz$^{2}$}
\affiliation{$^1$Institut de Physique Th\'eorique, CEA-Saclay, 91191 Gif-sur-Yvette, France\\
$^2$ International Institute of Physics, 
Universidade Federal do Rio Grande do Norte, 59078-400 Natal-RN, Brazil \\
$^3$Materials Science Division, Argonne National Laboratory, Argonne, IL  60439,  USA\\
$^4$ Condensed Matter Theory Group, CPMOH, UMR 5798, Universit\'e de Bordeaux I, 33405 Talence, 
France
}

\begin{abstract}
We argue that near a Kondo breakdown critical point, a spin liquid with spatial modulations can form.
Unlike its uniform counterpart, we find that this occurs via a second order phase transition.
The amount of entropy quenched when ordering is of the same magnitude as for  
an antiferromagnet.  Moreover, the two states are competitive, and at low temperatures
are separated by a first order phase transition.
The modulated spin liquid we find breaks $Z_4$ symmetry, as recently seen in the hidden
order phase of URu$_2$Si$_2$.
Based on this, we suggest that the
modulated spin liquid is a viable candidate for this unique phase of matter.
\end{abstract}

\maketitle
   
 The hidden order (HO) phase in URu$_2$Si$_2$ is a long standing problem in 
 condensed matter physics \cite{mydoshfirst},
and has recently received renewed attention from both experiment and theory.
This metal can be characterized as a moderate heavy fermion with a Sommerfeld coefficient  
of $\sim$ 180 mJ/mol K$^2$. 
  It undergoes a mean field-like second order phase transition at $T_0=17 K$, 
 characterized by a large jump in the specific heat. The amount of entropy
   quenched at the transition is substantial ($\approx$  1.38 J/mol K), which
   is of order of 24\% of the entropy of a local $f$ doublet. 
   Despite many years of intense investigation, the nature of the hidden order remains controversial.
    A number of theoretical propositions have been made,
   which can be divided into itinerant types, where the order parameter originates from
   delocalized $f$ electrons, and localized types, where it is believed that the
   local levels of a U 5f$^2$ ion are involved \cite{balatskyfirst,premi}. 
   
   
   Experimental data show evidence for both itinerant and localized character of the  order parameter.  Under pressure, the phase diagram evolves from the hidden order phase to an antiferromagnet (AF)  \cite{villaume}.   Recent inelastic neutron scattering (INS) measurements  show that in the HO phase, a resonance occurs at  a commensurate wave vector  $Q_0= (1,0,0)$ \cite{wiebe,bourdarot}, which transforms  into a strong elastic AF signal  for pressures $P$ $\geq$ 5 kbar.
   An  inelastic resonance at $Q^*=(1\pm0.4,0,0)$ occurs in both the HO and AF phases.
    The compensated nature of this metal leads to several quasi-nested portions of the  Fermi surface, which could account for the $Q$ vectors of these resonances, as well as the formation of an SDW \cite{elgazzar,janik}.  Recent STM experiments \cite{yazdani,davis} reveal the opening of a gap in the $dI/dV$ characteristic at $\approx T_0$  inside the HO phase. Moreover, careful analyses  of  specific
heat \cite{balatskyfirst} as  well as thermal transport \cite{kamran} have revealed a strong analogy with the superconductor CeCoIn$_5$. This body of observations, combined with the opening of a Fermi surface gap inferred from conductivity \cite{maple,jeffries}, angle resolved photoemission \cite{andres}, infrared spectroscopy \cite{bonn,vdmarel}, and the Hall effect \cite{oh}, constitute evidence for an itinerant mechanism.
   The localized viewpoint is based on the observation that under pressure, the first order transition between the HO and AF phases shows no distinct changes in the transport properties. The resistivity, for
 example, is continuous through the transition \cite{hassinger}. From this observation, one can expect  the HO phase to have strong similarities with the AF phase,
in particular a doubling of the unit cell.
      
    In this paper, we offer a new idea for  explaining the mysterious hidden order, which  
    naturally interpolates between the itinerant and localized viewpoints. The idea amounts to the observation that a  spin liquid  with spatial modulations that break translational symmetry can form
via a second order transition with a large jump in the specific heat, while remaining  ``hidden'' to most experimental probes.  
     
     The concept of a spin liquid  dates from the early work of Fazekas and Anderson that a
      resonating valence bond (RVB) state might describe frustrated spin systems on a 
      triangular lattice \cite{fazekas}.  Subsequent work by Anderson extended this concept to
      describe high temperature cuprate superconductors \cite{anderson87}.
     In a seminal paper \cite{bza}, the uniform spin liquid was understood as being a superposition of bonding and anti-bonding valence bond singlets, stabilized by quantum fluctuations.
    This concept has re-emerged recently in the context of quantum critical points (QCP) in
    heavy fermions  \cite{burdin,senthil,paul}.  The common understanding is  that the anomalous properties of these metals arise from a competition between the formation of magnetic singlets between localized spins, and the formation of Kondo singlets.  Recently, however, a  body of experimental observations 
     combined with theoretical insights has concluded that the formation of magnetic singlets is most probably the dominant mechanism for quenching the entropy of the local $f$ spins \cite{pines}. The complexity of actinide ions, under the combined influence of  spin-orbit, Hunds rules, and geometric frustration, create the optimal grounds  to favor the formation of valence bond singlets that eventually quench the high temperature entropy . A spin  liquid can simply be re-defined as a regime of localized spins in which the entropy is quenched despite the absence of long range order.
     
     In the RVB formalism, the spin liquid can be described within the $t-J$ model \cite{bza}, which in the limit of half filling reduces to the Heisenberg model, with fermionic spin operators
     \beq \label{eqn1}
     H_0=J\sum_{\langle i, j \rangle, \sigma\sigma'}  \chi^\dagger_{i \sigma } \chi_{i \sigma'} \chi^\dagger_{j \sigma'} \chi_{j \sigma} \eeq%
     where $ \chi_{i \sigma}^\dagger$ ($\chi_{i \sigma}$) are creation (annihilation) operators for fermions with spin 1/2, charge 
     zero, and gauge charge +e, commonly dubbed `spinons'. For simplicity, we restrict ourselves to the square lattice in two dimensions,
leaving a three dimensional generalization more appropriate for URu$_2$Si$_2$ to later work.
Here $\sigma=\pm$  is the spin index of the SU(2)  representation,  and spinons are subject to the constraint of one per site: $\sum_\sigma \chi^\dagger_{i \sigma} \chi_{i \sigma } = 1$. In the mean field approximation, the interaction term in (\ref{eqn1}) can  be decoupled in a variety of effective fields to minimize the free energy, including valence bond singlets, 
     $\sum_\sigma \langle \chi^\dagger_{i \sigma} \chi_{j \sigma} \rangle $, valence bond ``pairs'',  
     $\langle \chi^\dagger_{i \sigma} \chi_{j \bar \sigma} - \chi^\dagger_{i \bar \sigma} \chi_{j \sigma} \rangle $, or AF order, 
     $\sum_{\alpha \beta }\langle  \chi^\dagger_{i \alpha} {\bf \sigma}_{\alpha \beta } \chi_{i \beta} e^{i {\bf Q} \cdot {\bf r}_i}\rangle $,
     with $\bf \sigma$ the SU(2) spin matrix, and $Q$ the AF wave vector.

\begin{figure}[tbp]
\centerline{\includegraphics[width=3.4in]{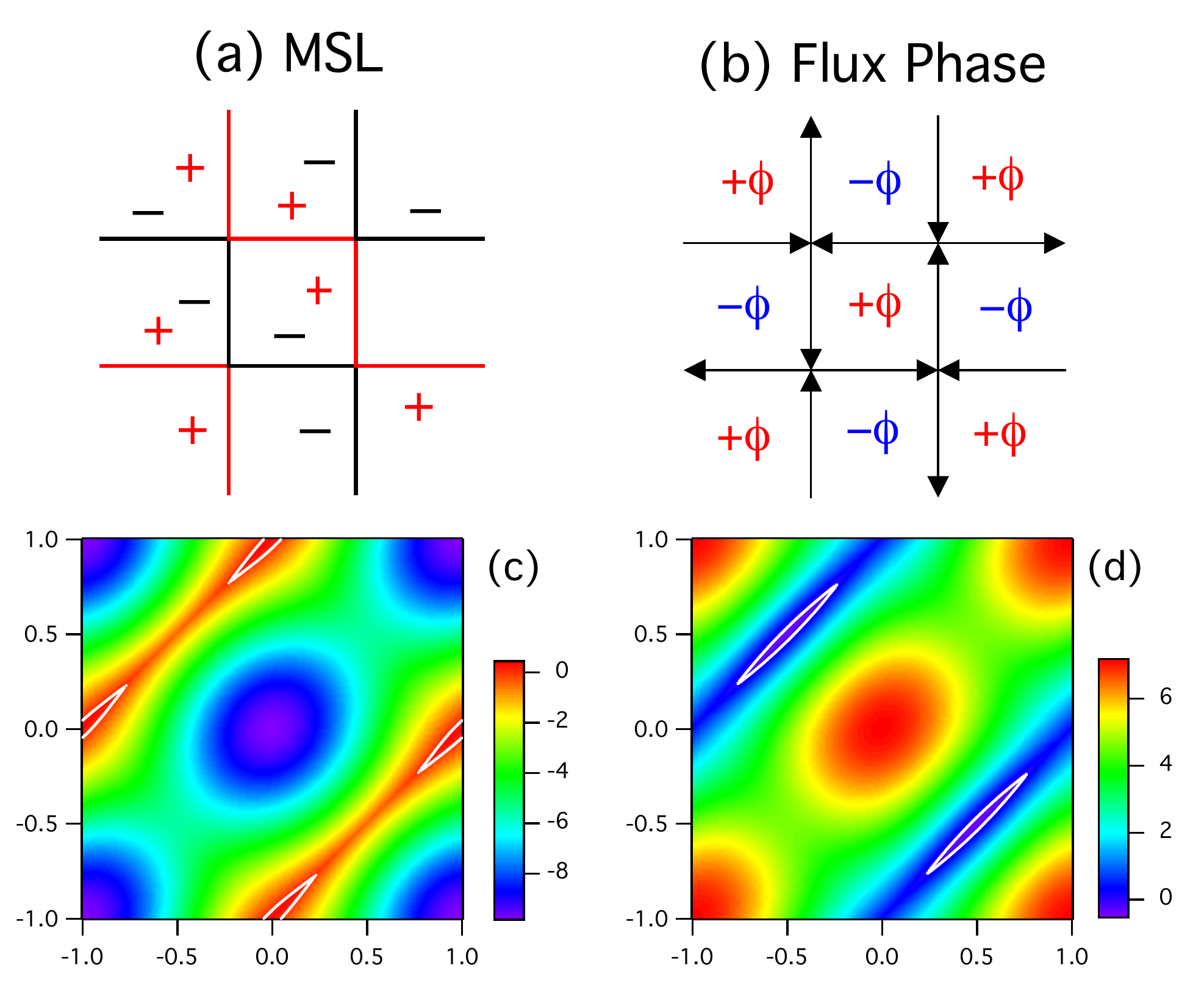}}
    \caption{(Color online) the MSL phase  (a) and the  $\pi$-flux phase (b). Note the orientation of the bonds in the latter case.
(c,d): the spinon dispersions for the MSL (bands 1 and 2).
 Note the breaking of $Z_4$ symmetry, and the small hole pockets for band 1 and
the small electron pockets for band 2 (white curves).
 The parameters are $t'$=0.1 and $J_{SL}$=5, with $\phi_0$=2.11 and $\phi_Q$=1.27.}
\label{fig1}
\end{figure}

   In the  RVB theory for cuprate superconductors, uniform valence bond order parameters 
   $\langle \chi^\dagger_{i \sigma} \chi_{j \sigma} \rangle =  \varphi_0 \delta_{{\bf r}_i, {\bf r}_j + z}$ where $z$ is the index of nearest or next nearest neighbors, or flux phases $ \langle \chi^\dagger_{i \sigma} \chi_{j \sigma} \rangle = \varphi_0e^{i \Phi /4 n_{ij}}$ 
   where $\Phi$ is the flux per plaquette and $n_{ij} = (r_i - r_j)/a$ (with $a$ the lattice spacing) is a bond  orienting number, are commonly introduced. In the $\pi$-flux phase, for example, when a spinon cycles around a plaquette, the circulation of the phase of the order parameter generates a magnetic flux of magnitude $\pi$ (Fig.~\ref{fig1}).
   Here we introduce another kind of valence bond order parameter, with real space modulations of the bond centers, which we denote as the modulated spin liquid (MSL):
   \beq \label{eqn2}
    \sum_\sigma \langle \chi^\dagger_{i \sigma} \chi_{j \sigma} \rangle=  \delta_{i, j + z} 
    \left [ \phi_0 + \frac{\phi_{\bf Q}}{2}\sum_{\pm}
    \ e^{\pm i[\theta + {\bf Q }\cdot \left ( {\bf r}_i + {\bf r}_j \right ) /2]}  \right ] \eeq%
     We see that the value of the bond acquires an oscillating sign from site to site. Since the phase on the bond is not oriented,  no flux is generated when a spinon cycles around a plaquette.
   This order parameter doubles the unit cell associated with the dual lattice of bonds, 
   breaking the $Z_4$ symmetry of the underlying square lattice (Fig.~\ref{fig1}). Whereas the flux phase is typical of lattices in two dimensions, the MSL can easily be generalized to three dimensions.
   
   The interaction term in (\ref{eqn1}) can be decoupled
   as either a spin liquid (SL) or an AF. For given $J \equiv J_{SL}+J_{AF}$ we consider $J_{SL}$
   and $J_{AF}$ as tuning parameters in our model, which thus acquire a phenomenological character.
   Considering that the spin liquid can have both a uniform  and a modulated component,
   the corresponding decouplings are performed via Hubbard-Stratonovich transformations on each bond $ij$, 
     leading to the following Lagrangian
           \bea  \label{eqn3bis}
     &&\hskip-.25cm {\cal L}_{0}   =  \sum_{i\sigma}\chi^\dagger_{i \sigma} \left( \partial_\tau  +\lambda_{i}
      +\sigma\sum_{z} m_{i+z}\right)  \chi_{i \sigma} -  \sum_i \lambda_i  \\
      &&\hskip-.4cm +\sum_{\langle i, j \rangle, \sigma} \left[ \varphi_{i j} \chi^\dagger_{i \sigma} \chi_{j\sigma}  
      +c.c.\right]
       +\sum_{\langle i, j \rangle } \left [ \frac{1}{J_{SL}}   \vert\varphi_{ij}\vert^2  
       -  \frac{1}{2 J_{AF} }  { m}_i{m}_j \right ] \nonumber \eea%
       Here $\varphi_{i j}$ is the Hubbard-Stratonovich field introduced for SL decoupling of the bond $ij$, and 
       ${m}_{i}$ arises from the AF decoupling of the site $i$. 
       In the following, these will be replaced by their constant, self-consistent, mean-field expressions, 
       $\varphi_{ij}=-J_{SL}\sum_{\sigma}\langle \chi^\dagger_{i \sigma}\chi_{j \sigma}\rangle$, and 
       ${m}_{i}=J_{AF}\sum_{\sigma}\sigma\langle \chi^\dagger_{i \sigma}\chi_{i \sigma}\rangle$. Note that the SL field is defined 
       on the dual lattice, whilst the AF one is defined on the initial (square) lattice. In general, $\varphi_{ij}$ can have a non-zero 
       imaginary part,  which is the case, for example, in a $\pi$-flux phase. 
       Here, we consider real SL fields only, which reflects a symmetry $\varphi_{ij}=\varphi_{ji}$ on each bond. Such a symmetry 
       being incompatible with the occurrence of a magnetic flux, the resulting modulated SL phase is necessarily of a 
       different nature. Note that in two dimensions, flux phases might still coexist with the present MSL phase, but the study of 
       this phenomenon is beyond the purpose of the present work. 
       We introduce the Fourier transformed fields, $\varphi_{\bf q}$ and $m_{\bf q}$. 
       Our analysis concentrates onto the three following mean-field parameters: the uniform SL, 
       $\phi_{0}\equiv \varphi_{(0,0)}$, the modulated SL, $\phi_{\bf Q}\equiv \varphi_{(\pi,\pi)}$, and the N\'eel AF, 
       $S_{\bf Q}\equiv m_{(\pi, \pi)}$,
noting that $(\pi,\pi)$ has a lower free energy than $(\pi,0)$.

     An intuitive description of the MSL phase, compared to other possible SL phases, can be obtained from 
     finite size versions of our model.  For two sites and $J_{AF}=0$, 
     the effective mean-field Hamiltonian~(\ref{eqn3bis})
     can be decomposed into bonding and anti-bonding eigenmodes, 
     $H_{12}= \frac{\vert\varphi_{12}\vert^{2}}{J_{SL}}+\varphi_{12}\sum_{\sigma}\left ( 
     \chi^\dagger_{A \sigma}\chi_{A \sigma}-\chi^\dagger_{B \sigma}\chi_{B \sigma}\right )$, 
     with $\chi_{B\sigma}\equiv \frac{1}{\sqrt{2}}\left( \chi_{1 \sigma} +\chi_{2 \sigma}\right)$ and 
     $\chi_{A\sigma}\equiv \frac{1}{\sqrt{2}}\left( \chi_{1 \sigma} -\chi_{2 \sigma}\right)$. These 
     modes are reminiscent of the singlet,
     and the $S_z$=0 triplet, which diagonalize the initial, two sites, Hamiltonian~(\ref{eqn1}). 
     Invoking the saddle-point self-consistent relation for $\varphi_{12}$, we find a ground state energy, 
     $\langle H_{12}\rangle =-\frac{\vert\varphi_{12}\vert^{2}}{J_{SL}}$. 
     The $\pm\varphi_{12}$ degeneracy reflects the $U(1)$ local gauge symmetry of the Hamiltonian~(\ref{eqn1}), which is
      invariant with respect to the transformation $\chi_{2\sigma}^{\dagger}\to -\chi_{2\sigma}^{\dagger}$.  
     The groundstate of the mean-field effective model can therefore arbitrarily be chosen as a bonding or an anti-bonding mode. 
     Due to this gauge symmetry, for the two sites model, all SL mean-fields are equivalent and mimic the energy splitting between 
     a singlet ground state and one of the triplet excited states. This equivalence does not hold anymore for bigger systems, where the 
     number of sites (i.e., the number of local gauge symmetries) becomes smaller than the number of bonds. 
     In the MSL case on the square lattice, however, the bonding and anti-bonding character oscillates from site to site, 
     since the sign of the hopping parameter, $\varphi_{ij}$, is oscillating. \\
     
     The MSL can be considered as the true RVB parent of the AF phase.  It can also be viewed as a liquid phase of dimers, where a resonance moves  between different dimer coverings of the lattice. In  three dimensions, the liquid phase of dimers can coexist with the breaking of translational invariance.

Fig.~\ref{fig1} depicts the spinon dispersion of the MSL. The breaking of $Z_4$ 
symmetry is obvious. Comparison of the free energy and specific heat jump at the transition reveals that the MSL and the AF decouplings  are almost degenerate at the mean field level. The MSL can thus be considered as the RVB parent of the AF order.
     
      The issue of whether a gauge invariant Lagrangian develops crossovers or phase transitions between the Higgs, confined and Coulomb phases is an old one and much related to the presence of  instantons in the system \cite{anderson}.  In the case of the MSL, we argue that  the breaking of $Z_4$ symmetry is enough to ensure a second order transition \cite{chandra-larkin}.  This is to be contrasted with the $\pi$-flux phase or the   possible condensation of the holons with modulations $b_i e^{i {\bf Q}\cdot {\bf r}_i}$ \cite{paul}.  In both of those cases,  the order parameter would be sensitive to the effect of gauge fluctuations, and the presence of instantons at finite temperature  generates a crossover. 

      \begin{figure}[tbp]
\centerline{\includegraphics[width=3.4in]{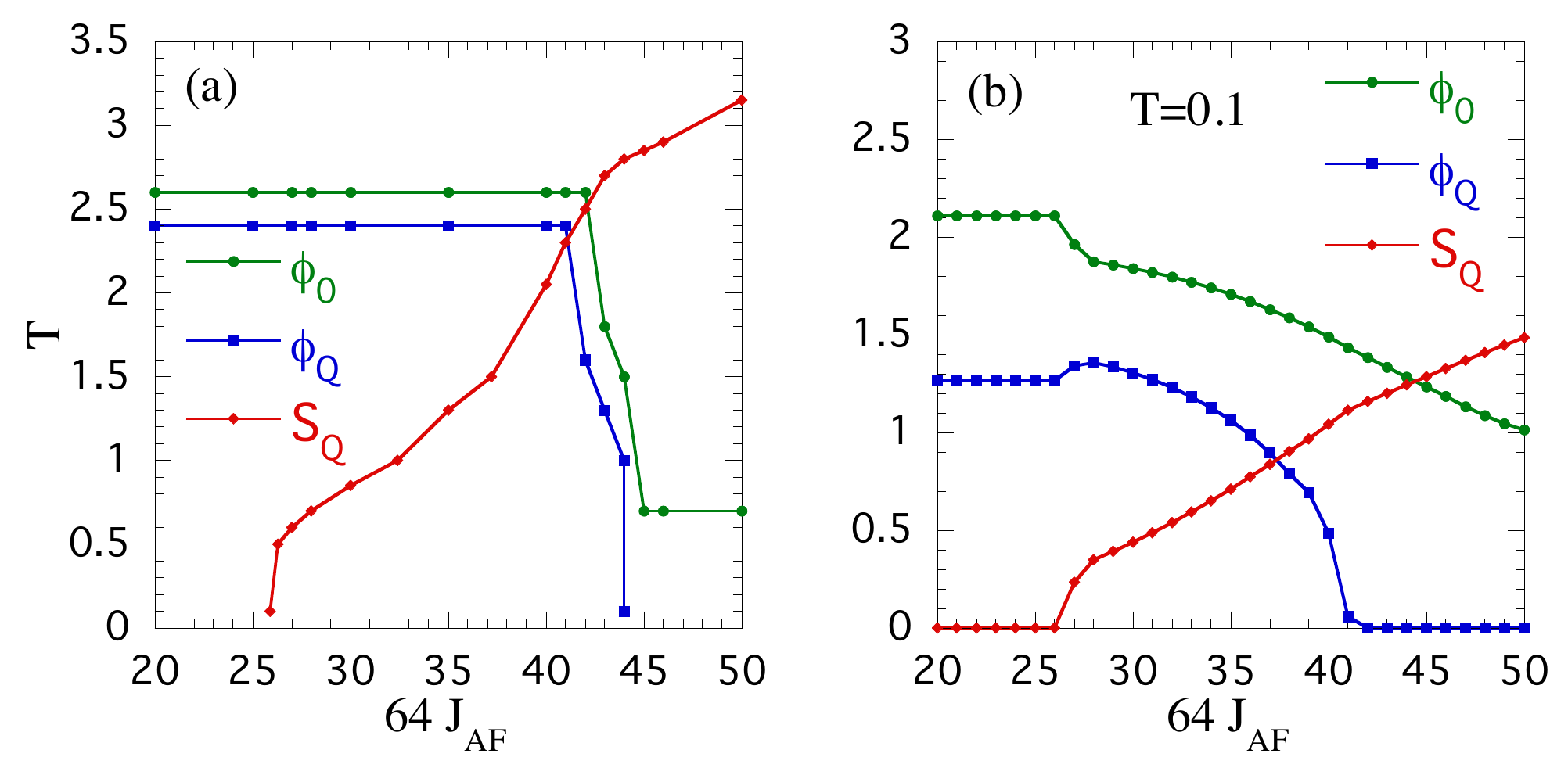}}
    \caption{(Color online) (a) mean field phase diagram in the $(T,J_{AF})$ plane from (\ref{free}) for a 
    square lattice model with $t'=0.1$ and $J_{SL}$=5.
    Below the green curve (circles), $\phi_0$ turns on and below the blue one (squares), 
 $\phi_Q$ condenses. $S_Q$ condenses below the red curve (diamonds).
    For low $T$ this is a first order transition, at higher $T$ a second order one.
     (b) Variation of  $\phi_0$, $\phi_Q$, and $S_Q$ as a function of $J_{AF}$ 
     for $T$=0.1.}
\label{fig2}
\end{figure}

In the remainder of the paper, we discuss the compatibility of the MSL with the HO phase in  URu$_2$Si$_2$, and its relation to the AF phase.
To proceed, we solve a simpler model where the conduction electrons are ignored,
and the f electrons are treated as a single orbital on a square lattice
with $Q=(\pi,\pi)$.  The resulting
free energy per site can be written as:
\beq \label{free}
F = -\frac{ k_B T}{N} \sum_{k,i}\ln(1+e^{-\beta\omega_i(k)}) + \frac{2\left (\phi_0^2+\phi_Q^2 \right )  }{J_{SL}}
+ \frac{S_Q^2}{J_{AF}} 
\eeq
where $\beta=1/(k_BT)$, $N$ is the number of sites,
 $i$ runs over the two  bands (each of which are Kramers degenerate),
and the spinon dispersions $\omega_i(k)$ are given by
\bea
\omega_i(k) & = & \frac{\phi_0}{2}(\epsilon_k+\epsilon_{k+Q})-\mu_f \\
& & \pm \sqrt{\frac{\phi_0^2}{4}(\epsilon_k-\epsilon_{k+Q})^2+(\phi_Q\epsilon^{NN}_{k-Q/2})^2+16 S_Q^2} \nonumber
\eea
with
\beq
\epsilon_k = -2(\cos(k_xa)+\cos(k_ya))-4t'\cos(k_xa)\cos(k_ya)
\eeq
and $\mu_f$ the chemical potential of the $f$ electrons adjusted so that the system  is at half filling.
 Here, $t'$ represents the next near neighbor contribution to $\phi_0$.  For $\phi_Q$, only
 near neighbor bonds are considered (with $\epsilon^{NN}$ the first term of the previous
 equation), and to obtain a real field, the quantity $\theta$ in (\ref{eqn2}) is set to $\pi/2$.
 We minimize the free energy using Powell's method, and use a root finder to 
 determine $\mu_f$ for each choice of $(\phi_0, \phi_Q, S_Q)$ \cite{NumRec}.
 
In Fig.~\ref{fig1}, we show the spinon dispersions for the MSL case.
Note the pronounced breaking of $Z_4$ symmetry, which has been recently detected in the HO
phase from susceptibility measurements \cite{matsuda}.
A large part of the Fermi surface (as defined when $\phi_Q$ is zero)
is gapped upon ordering  \cite{mydoshfirst}, which gives a natural explanation for the amount of entropy quenched, and is consistent with the  Hall  \cite{oh}, thermal conductivity \cite{kamran} and quantum
oscillation \cite{nakashima} data that suggest
that 90\% of the carriers disappear  at the transition. The dispersion for the AF phase is very similar,
except in our simplistic approximation, a full energy gap occurs once $S_Q > t'\phi_0/2$.
We also find that the the order parameters in each phase have similar magnitude.
One can tune between the two phases by varying $J_{AF}$ relative to $J_{SL}$, as shown in
Fig.~\ref{fig2}.  Note that $\phi_0$ and $\phi_Q$ are quasi-degenerate and condense at almost the 
same $T$.  They would be equal if $t'=0$.  Above their condensation temperature, the free energy
goes as $-2k_BT\ln(2)$, which is just the free energy for a local $f$ doublet.
$S_Q$ condenses as $J_{AF}$ increases, via a first
order transition for low $T$, changing to a second order transition at higher $T$.  In the AF
phase, all three order parameters are at first non-zero, with $\phi_Q$ eventually disappearing
(at low $T$, we find a finite $\phi_0$).  The competition between the modulated component 
($\phi_Q$) and the AF order ($S_Q$) that gives rise to the first order behavior is obvious from 
Fig.~\ref{fig2}.
This reproduces the qualitative features of the  experimental phase diagram under pressure where
a first order transition occurs between the HO and AF phases.
Note that since both orders double the 
unit cell, no noticeable feature is expected to be seen in the electrical conductivity at this 
transition.
    
From the more complete theory, we expect that the MSL phase will be stabilized near a Kondo breakdown QCP \cite{paul}, reflecting the localization of the two 5f electrons per U site
due to strong Coulomb forces. At the localization transition, the two 5f electrons decouple from the conduction electrons to  form a spin liquid.  The effective hybridization is renormalized to zero  at the QCP.  The tunneling experiments confirm the opening of a hybridization gap  at $\approx T_0$ between heavy  and light hole-like bands \cite{yazdani,davis}.  We interpret  the hybridization gap as the effective hybridization between the spinons and the conduction electrons.  Note that the spinons are very difficult to observe; they can be detected by transport only when they hybridize with the conduction electrons.  This could explain why the  heavy band seems to disappear above $T_0$     
    
    INS experiments show an inelastic resonance at  a commensurate wave vector $Q=(1,0,0)$ in the HO phase \cite{bourdarot}, which becomes an elastic peak  in the AF phase.
Our theory also has an inelastic resonance at $Q_{AF}$ since the AF state has a higher free energy
in the HO phase.  We speculate that the incommensurate peaks are due to nesting of the spinon Fermi surface.

        In conclusion, the modulated spin liquid is an interesting phase of matter which has many properties compatible  with the hidden order phase in URu$_2$Si$_2$.  The hidden nature of our order parameter simply relies on the fact that a spin liquid is hardly detectable if no long range symmetry is broken. We believe that it is a viable candidate for the solution of   this long standing mystery.
          
 CP acknowledges an ICAM travel fellowship, and  the Aspen Center for  Physics where this work was started.  MN was supported by the U. S. DOE, Office of Science, under contract DE-AC02-06CH11357.
CP and SB acknowledges the financial support of the Brazilian Ministry of  Sciences and Technology and the hospitality  of the IIP, Natal, where part of this work was completed. AF is supported by the CNPq and by the BMST.

\end{document}